\documentclass[letter,12pt]{article}
\pdfoutput=1
\usepackage{jheppub} 
\usepackage{graphicx,amssymb,amsmath,color,slashed}
\usepackage[T1]{fontenc}

\def\beq{\begin{equation}}
\def\eeq{\end{equation}}
\def\ceq{\end{equation} \begin{equation}}
\def\bea{\begin{eqnarray}}
\def\eea{\end{eqnarray}}
\def\bei{\begin{itemize}}
\def\eei{\end{itemize}}
\def\bmat{\begin{matrix}}
\def\emat{\end{matrix}}
\def\ble{\begin{flushleft}}
\def\ele{\end{flushleft}}
\def\={\,=\,}
\def\+{\,+\,}
\def\-{\,-\,}

\def\TeV{\,{\rm TeV}}

\def\SU2{{\rm SU}(2)}

\def\axi{\widetilde{a}}

\newcommand{\Fig}[1]{Fig.~\ref{#1}}
\newcommand{\Eq}[1]{Eq.(\ref{#1})}
\newcommand{\Sec}[1]{Sec.~\ref{#1}}

\begin{document}

\title{Resolving the existence of Higgsinos in the LHC inverse problem}
\author{Sunghoon Jung} 
\affiliation{School of physics, Korea Institute for Advanced Study, Seoul, Korea}
\abstract{The LHC inverse problem is infamously challenging when neutralinos and charginos are heavy and pure and other superparticles are decoupled. This limit is becoming more relevant to particle physics nowadays. Fortunately, in this limit, Higgsinos produce a distinctive signature if they are the LSPs or NLSPs. The identifying signature is the presence of equal numbers of $Z$ bosons and Higgs bosons in NLSP productions and subsequent decays at hadron colliders. The signature is derived from the Goldstone equivalence theorem by which partial widths into $Z$ and Higgs bosons are inherently related and from the fact that Higgsinos consist of two \emph{indistinguishable} neutralinos. Thus it is valid in general for many supersymmetry models; exceptions may happen when Higgsino NLSPs decay to weakly coupled LSPs such as axinos or gravitinos.}

\preprint{KIAS-P14020}

\emailAdd{nejsh21@gmail.com}
\maketitle

%%%%%%%%%%%
\section{Introduction}

The absence of supersymmetry(SUSY) signatures at the LHC pushes us into the regime of ${\cal O}(100-1000)$GeV inos\footnote{We write ``inos'' for any neutralinos and charginos in the introduction but will refer only to electroweak-gauginos, Higgsinos and possibly gravitinos and axinos afterwards.} and heavy-decoupled scalar superparticles. Such heavy inos are typically well-separated in mass leading to minimal ino mixings and degenerate charginos and neutralinos of the same kind. Hereafter, such parameter space is called ``the split limit''. The split limit is not only phenomenologically supported, but can also be theoretically motivated as charged SUSY breaking -- SUSY breaking without singlets -- generically leads to it~\cite{Randall:1998uk,Wells:2003tf}. 

In the split limit, the LHC inverse problem~\cite{ArkaniHamed:2005px} arises and is difficult to resolve. The problem can be described as: (1) first of all, the discovery of gauginos and Higgsinos is difficult, and (2) the extraction of model paramters, foremost importantly the identities and the masses of inos, is often subject to multiple interpretations. 

The inverse problem has been addressed in more general SUSY parameter space including light sfermions and gluinos by carrying out a scan over a huge parameter space and a dedicated collider simulation and by considering a huge set of collider and astrophysical observables~\cite{Allanach:2007qk,Berger:2007yu}, or by adapting a sophisticated statistical analysis~\cite{Balazs:2009it}. These works have provided an unprecedented amount of useful information on the multi-dimensional parameter space of SUSY. The split limit, however, has much fewer parameters and particles relevant to collider analysis allowing us to more analytically approach the problem. It is a well-motivated and meaningful subset which deserves the detailed study on its own.

Before we get into the analysis, we would like to illustrate how the problem arises in the split limit. In the general parameter space that has been studied widely, a change in the ino sector can be accompanied by suitable changes of other sectors such as slepton masses and mixing angles so as to yield the same observables~\cite{ArkaniHamed:2005px}. Does the split limit also have enough parameters to induce degeneracies in the data interpretation? 

\begin{figure}[t] \centering
\includegraphics[width=0.32\textwidth]{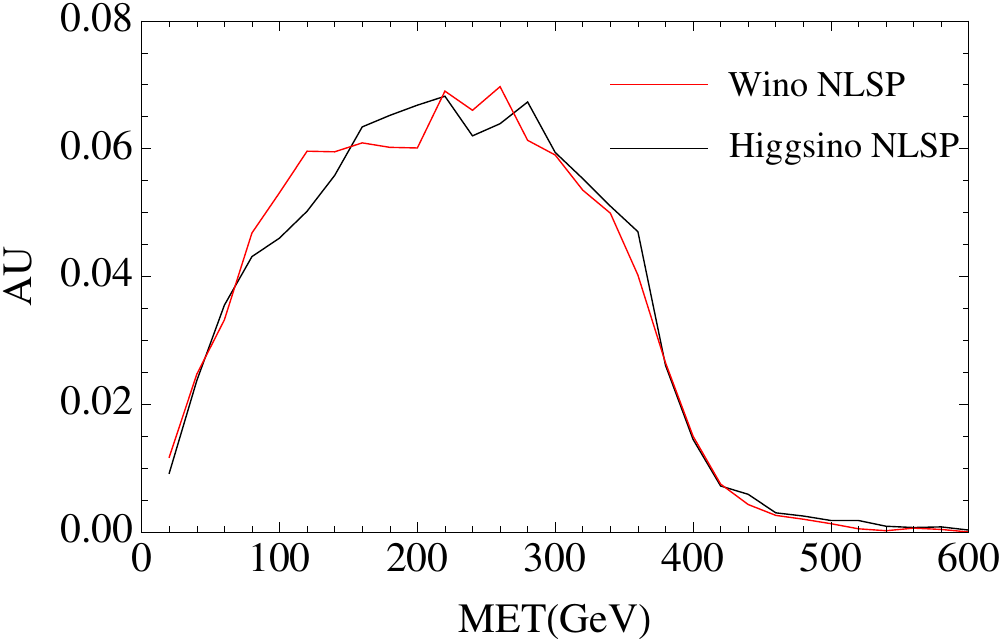}
\includegraphics[width=0.32\textwidth]{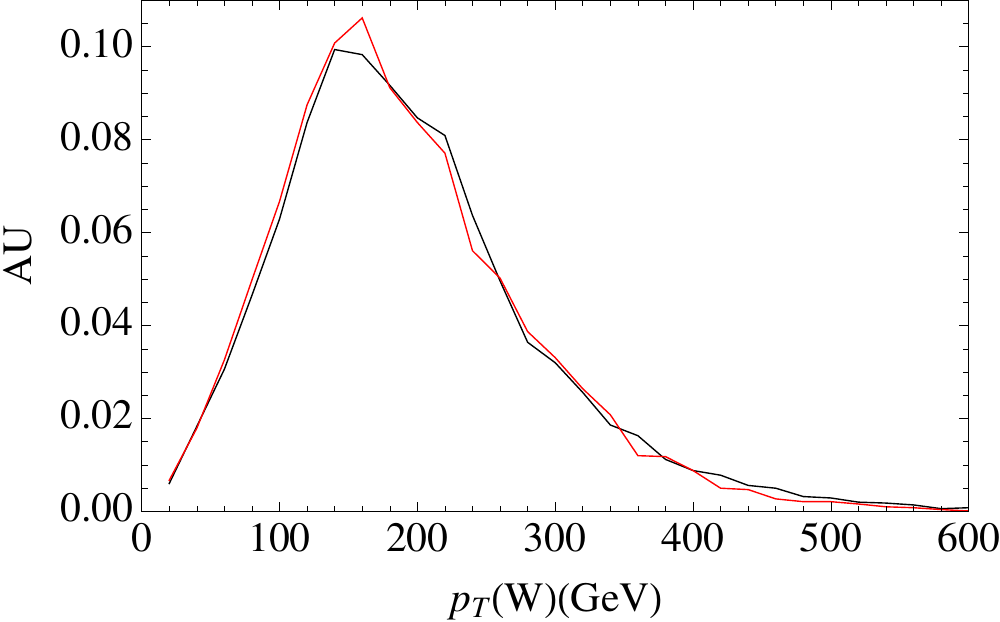}
\includegraphics[width=0.32\textwidth]{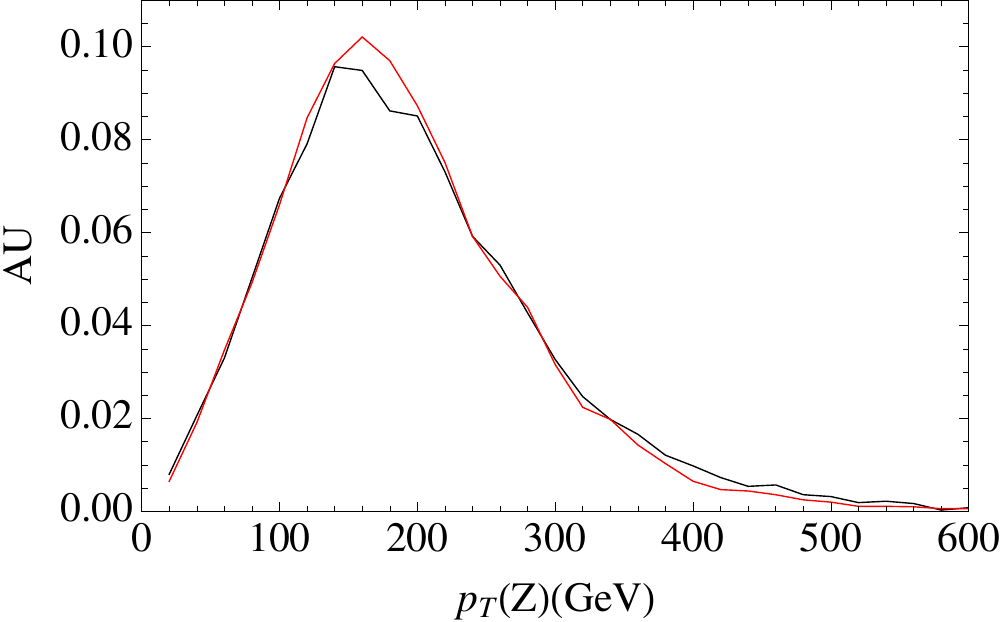}
\caption{Illustrating the LHC inverse problem in the split limit. Differential cross-sections of MET(left), $p_T(W)$(mid) and $p_T(Z)$(right) from the decays of NLSP pairs, $\chi^+ \chi^0$ as an example, are compared between the Wino-NLSP(red) and the Higgsino-NLSP(black) models with the Bino-LSP. Actually, not only the identities of the NLSP, but also many parameters in the models are different; however, these easiest observables may not distinguish the models. Models are defined in Table.\ref{tab:inverse} and more discussions are in text.}
\label{fig:inverse}
\end{figure}

We create two degenerate models in the split limit for illustration in Table~\ref{tab:inverse}.  They are degenerate at least in the early  stage of the LHC14 running because they produce same counting and differential observables accessible in the early stage. The easiest channels to search for the electroweakino sector will be di-vector-boson channels leading to multi-lepton and missing transverse energy(MET) signatures. Three diboson channels, $WW,WZ$ and $ZZ$, lead to mutually different multi-lepton signals, so the rate of each category is an observable in the inverse map; see Table~\ref{tab:inverse}. In addition to these lepton countings, the differential distributions of the $p_T$ of $W,Z$ bosons and MET are most important ones affecting many other observables. We compare these representative distributions for illustration in \Fig{fig:inverse}. As shown in Table~\ref{tab:inverse} and \Fig{fig:inverse}, not only total rates of each diboson category, but also the shapes of those distributions are almost identical in the two models. Here we note that contributions from all possible pair productions of nearly degenerate \emph{indistinguishable} next-to-lightest inos(NLSP) are added -- these are what we will observe at collider.

\begin{table}[t] \centering
\begin{tabular}{c |c || c | c | c}
\hline \hline
Model & parameters ($M_1, M_2, \mu, t_\beta$) & $\sigma(W^+W^-)$  & $\sigma(W^\pm Z)$  & $\sigma(ZZ)$ \\ 
\hline \hline
Wino-NLSP & $0.5\TeV, 1.0\TeV, -2.0\TeV, \, 4.3$ & 0.60 fb & 1.1 fb & 0 fb \\
\hline
Higgsino-NLSP & $0.2\TeV, 2.0\TeV, 0.8\TeV, \, 2.0$ & 0.61 fb & 1.1 fb & 0.02 fb \\
\hline \hline
\end{tabular}
\caption{Definitions of the models used in the illustration in \Fig{fig:inverse}. Also shown are diboson production cross-sections from all possible NLSP pair productions and their subsequent decays. All other superparticles are heavier. LHC14. See text for how we create these models.}
\label{tab:inverse} \end{table}

The problem is that two models are quite different. Most importantly, the identities and  the masses of the NLSPs are different. For any given identity of the NLSP, we were able to tune the NLSP mass for production rates, $t_\beta$ for branching ratios and the mass gap between the LSP and the NLSP for the shapes of spectra. This kind of degeneracy generally exists in the split limit. 

In this paper, we consider the split limit and study an \emph{observable} relation of the NLSP decay that can confidently tell us the existence of Higgsinos as either LSPs or NLSPs.  We have already made one of our main observations that all indistinguishable decay processes should be added up to produce an observable. We will see that this makes the Higgsino unique.

Our basic study model consists of electroweakinos and Higgsinos (with heavy-decoupled gluinos). But we also consider axinos or gravitinos additionally. Additional considerations will further support our claim and help to generalize our arguments. Exceptions to the claim will also be found in such cases with weakly interacting LSPs.

We will call all nearly degenerate (next-to-)lightest states by the (N)LSP -- we do not use the terminology such as co-NLSP. For example, NLSP Higgsinos refer to all nearly degenerate two neturalinos and one chargino stemming from $\widetilde{H}_u$ and $\widetilde{H}_d$ which are all heavier than lightest group of states; likewise, LSP Winos, e.g., refer to nearly degenerate charged and neutral Winos which are lightest states. We will consider only NLSP pair productions and subsequent decays since the inos are well-separated in mass and the production of NNLSP will be rarer. Otherwise, many more useful observables maybe constructable. 

In \Sec{sec:prelim}, we introduce the usefulness and caveats related to the Goldstone equivalence theorem. We then present a formal discussions of our signal in \Sec{sec:formal}. In \Sec{sec:production} it is argued that Higgsino productions satisfy the necessary condition for the signal. We then define the signal collider observable and carry out a numerical study without any approximations, in order to demonstrate the validity of the formal discussion in \Sec{sec:rzh}. Then we conclude.

%%%%%%
\section{Preliminaries on the Goldstone equivalence theorem} \label{sec:prelim}

\begin{figure}[t] \centering
\includegraphics[width=0.99\textwidth]{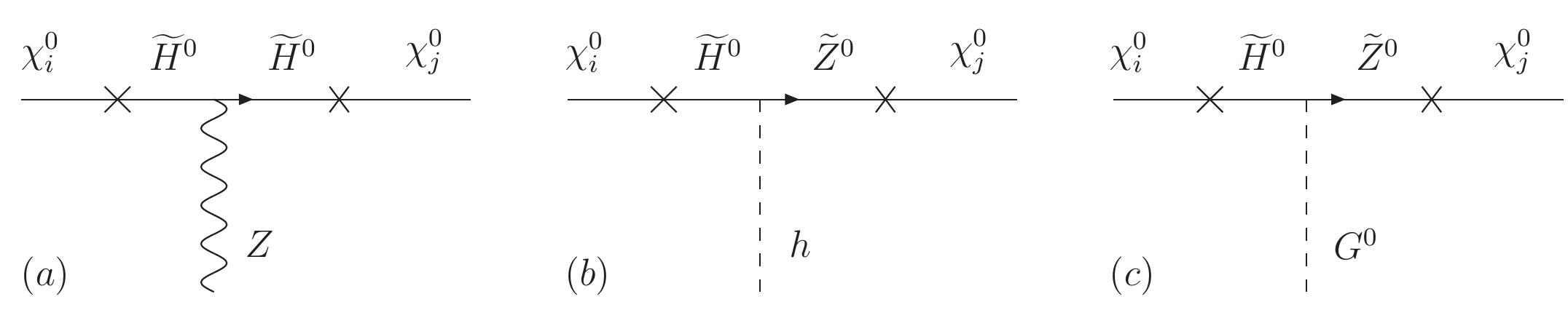}
\caption{Diagrams for generic processes $\chi_i^0 \to \chi_j^0 + Z,h,G^0$. Interactions of mass eigenstates can be approximately understood from their interaction-eigenstate mixtures and original interactions. In these diagrams, intermediate Higgsinos and Zinos are interaction eigenstates and crosses imply their mixtures in external mass eigenstates. These diagrams can provide useful estimations of the processes but should be used with some care as discussed in text.}
\label{fig:mixingdia}
\end{figure}

The Goldstone equivalence theorem~\cite{Cornwall:1974km,Lee:1977eg} says that the amplitude for the production of longitudinally-polarized energetic $Z$ bosons is equivalent to that of the Goldstone boson.

The theorem is useful in our study because neturalinos are heavy and well-separated in mass so that decays between them produce energetic $Z$ bosons. It is especially powerful when discussing Higgsino's interactions because Higgsinos directly couple to both $Z$ and $h$ bosons and necessarily to the Goldstone; thus, the processes of $Z$ and $h$ bosons are \emph{inherently} related in general -- we will be relating partial widths into $Z$ and $h$ in this paper. 

Let us consider two generic processes $\chi_i^0 \to \chi_j^0 + h,Z$ with $i \ne j$. One may compare two rates based on the first two diagrams in \Fig{fig:mixingdia}. If $\chi_i^0$ is Higgsino-like and $\chi_j^0$ is Zino-like\footnote{The Zino is a Bino-neutral Wino mixture whose mixing angle is precisely the weak mixing angle. Zinos inherit the coupling structure of $Z$ bosons, thus it is used to understand possible decay modes in \Fig{fig:mixingdia}. Zinos can even be mass eigenstates for some cases that we will discuss later. Photinos are orthogonal mixtures similarly defined in regard of photons.}, one may conclude that the decay into $h$ is more rapid because it does not need any small mixing insertions while the decay into $Z$ needs one small mixing insertion. However, this kind of arguments should be made with some care. When $m_i - m_j \gg m_Z$, the third diagram of decays into the Goldstone boson (hence, the longitudinal $Z$) becomes a good approximation of the first diagram. Indeed, the Goldstone diagram does not need any mixing insertion and can be expected to be comparable to the $h$ diagram. 

Technically, this happens as the growing longitudinal polarization vector of the $Z$ boson in energetic processes overcomes the small mixing insertion. A popular example of such Goldstone enhancement is the top decay; the decay rate into the longitudinal $W$ is enhanced by $m_t^2/m_W^2$ compared to the decay rate into the transverse components. In our ino study, we find the same Goldstone enhancements from $1/r_Z, 1/r_W \gg 1$ factors in \Eq{eq:dec-ch-Z}-\Eq{eq:dec-neu-W}. Indeed, the enhanced terms play a major role in approximating the full decay width into the $Z$ boson by the corresponding width into the Goldstone. We refer to Appendix \ref{sec:app-gold} for more discussions.

Throughout this paper, we will derive relations between partial widths into the $Z$ and $h$ using the Goldstone equivalence theorem. We will eventually obtain an interesting observable that can tell us the existence of the (light) Higgsinos.

%%%%%%%
\section{Partial width ratios into $Z$ vs. $h$} \label{sec:formal}

\subsection{In the presence of Higgsinos} 

In the split limit characterized by $m_i - m_j \gg m_Z$ and small neutralino mixings (and decoupled scalar superparticles), the NLSP neutralino decay width ratio is given by a ratio of scalar coupling squares (times a mild mass ratio factor)
\beq
\frac{\Gamma(\chi_i^0 \to \chi_j^0 Z)}{\Gamma(\chi_i^0 \to \chi_j^0 h)} \, \simeq \, \frac{ | D^{\prime L}_{Gij}|^2 \, (1-2\sqrt{r_j})}{ | D^{\prime L}_{hij}|^2 \, (1+2 \sqrt{r_j})},
\label{eq:neu-gold} \eeq
where $r_j = m_j^2/m_i^2$. The form of the simple ratio of scalar-gaugino couplings is implied by the Goldstone equivalence theorem. We refer to Appendix \ref{sec:app-gold} for the derivation, and we introduce notations and collect partial widths in Appendix \ref{sec:app-int}.

For the electroweakino-Higgsino case\footnote{The terminology means that either electroweak-gauginos are LSPs and Higgsinos are NLSPs or vice versa. The results we will obtain are independent on which one is LSP. We use similar terminology for other cases too.}, the ratio is more usefully expressed as
\beq
\frac{\Gamma(\chi_i^0 \to \chi_j^0 Z)}{\Gamma(\chi_i^0 \to \chi_j^0 h)} \, \simeq \, \frac{ |c_\beta N_{H_k 3} + s_\beta N_{H_k 4} |^2 \, ( 1-2 \sqrt{r_j} ) }{ | c_\beta N_{H_k 3} - s_\beta N_{H_k 4} |^2 \, ( 1+2 \sqrt{r_j} ) },
\label{eq:ew-higgsino} \eeq
where the index $H_k=i$ or $j$ indicates lighter($k=1$) or heavier($k=2$) Higgsino neutralinos. $H_k$ can be either decaying NLSP $i$ or LSP $j$; thus, this formula is valid for both cases of NLSP and LSP Higgsinos. We keep the terms $\pm2 \sqrt{r_j}$ that can be ${\cal O}(10\%)$ even if NLSP is 10 times heavier than LSP. Up to this size of correction, our formal discussion is valid; in \Sec{sec:rzh}, we will numerically demonstrate our formal discussions here.

In the split limit, the neutralino mass matrix becomes block-diagonal form; the Higgsino eigensystem is obtained from the following $2\times 2$ sub-matrix
\beq
\left( \bmat 0 & -\mu \\ -\mu & 0 \emat \right).
\label{eq:higgsino2by2} \eeq
The eigenvectors are $\chi_{H_{1,2}}^0 \simeq \frac{1}{\sqrt{2}}\left( \widetilde{H}_d^0 \pm \widetilde{H}_u^0 \right)$, and neutralino mixing matrix elements satisfy 
\beq
\frac{N_{H_1 3}}{N_{H_1 4}} \= - \frac{N_{H_2 3} }{N_{H_2 4}} \qquad \textrm{for both } \mu >0 \textrm{ and } \mu<0.
\label{eq:higgsinomixing}\eeq
This implies interesting relations of partial widths of two neutral Higgsinos. If the Higgsino is LSP (although exactly same arguments apply to the Higgsino NLSP case, we take the LSP example here for specific discussion and notational simplicity),
\beq
\Gamma(\chi_i^0 \to \chi_{H_1}^0 Z) \, \simeq \,  \Gamma(\chi_i^0 \to \chi_{H_2}^0 h), 
\ceq
\Gamma(\chi_i^0 \to \chi_{H_1}^0 h) \, \simeq \,  \Gamma(\chi_i^0 \to \chi_{H_2}^0 Z),
\ceq
\frac{\Gamma(\chi_i^0 \to \chi_{H_1}^0 Z)}{\Gamma(\chi_i^0 \to \chi_{H_1}^0 h)} \, \simeq \, \frac{\Gamma(\chi_i^0 \to \chi_{H_2}^0 h)}{\Gamma(\chi_i^0 \to \chi_{H_2}^0 Z)}.
\eeq
Since decay products into lighter and heavier Higgsinos are not distinguishable, what we observe is actually the sum of all decay products. The observable relation is then
\beq
\Gamma(\chi_i^0 \to \chi_{H_1}^0 Z) \ + \Gamma(\chi_i^0 \to \chi_{H_2}^0 Z) \, \simeq \, \Gamma(\chi_i^0 \to \chi_{H_1}^0 h) \+ \Gamma(\chi_i^0 \to \chi_{H_2}^0 h).
\label{eq:conc-lsp} \eeq
This means that we will observe the same numbers of $Z$ and $h$ bosons produced from NLSP decays. We express this statement as $R_{Z/h} \simeq 1$; the observable will be defined in \Sec{sec:rzh}. Related results are also discussed in Ref.~\cite{Han:2013kza}.

It is straightforward to repeat the same calculation for the NLSP Higgsino case and obtain the similar observable relation
\beq
\Gamma(\chi_{H_1}^0 \to \chi_j^0 Z) \ + \Gamma(\chi_{H_2}^0 \to \chi_j^0 Z) \, \simeq \, \Gamma(\chi_{H_1}^0 \to \chi_j^0 h) \+ \Gamma(\chi_{H_2}^0 \to \chi_j^0 h).
\label{eq:conc-nlsp} \eeq
This relation also means that we will observe the same numbers of $Z$ and $h$ bosons produced from NLSP decays; $R_{Z/h} \simeq 1$. In all, the same conclusion holds regardless of whether Higgsinos are LSPs or NLSPs.

Likewise, chargino decays are related by the Goldstone equivalence theorem
\beq
\frac{\Gamma(\chi_i^+ \to \chi_j^+ Z)}{\Gamma(\chi_i^+ \to \chi_j^+ h)} \, \simeq \, \frac{ |D^L_{Gij}|^2 + |D^L_{Gji}|^2 + 4Re[ D^L_{Gij} D^L_{Gji}] \sqrt{r_j} }{ |D^L_{hij}|^2 + |D^L_{hji}|^2 + 4Re[ D^L_{hij} D^L_{hji}] \sqrt{r_j} }.
\eeq
We again refer to Appendix~\ref{sec:app-gold} for the derivation and to Appendix \ref{sec:app-int} for notations. The decays between charginos can happen only for the Wino-Higgsino case. For the Wino-Higgsino case,
\beq
|D^L_{ij}|^2 + |D^L_{ji}|^2 + 4Re[ D^L_{ij} D^L_{ji}] \sqrt{r_j} \, \simeq \, \frac{1}{2}  \Big( \, |k_d|^2 + |k_u|^2 + 4Re[ k_d^* k_u^* ] \sqrt{r_j} \, \Big)
\eeq
giving
\beq
\frac{\Gamma(\chi_i^+ \to \chi_j^+ Z)}{\Gamma(\chi_i^+ \to \chi_j^+ h)} \, \simeq \, \frac{ c_\beta^2 + s_\beta^2 + 4 c_\beta s_\beta \sqrt{r_j} }{c_\beta^2 + s_\beta^2 + 4 c_\beta s_\beta \sqrt{r_j} } \=1.
\eeq
Thus, fortunately, each chargino decays equally to $Z$ and $h$ in the split limit -- it does not ruin the previously discussed neutral NLSP decay relation $R_{Z/h} \simeq 1$. The relation holds again regardless of whether Higgsinos are LSPs or NLSPs.

So far, we have considered the split limit. But the split limit is not strictly needed for the Goldstone equivalence theorem. One particularly relevant example is when Binos and Winos are nearly degenerate, i.e., $M_1 \simeq M_2$. This is not the split limit because Binos and Winos maximally mix. Nevertheless, since the Bino-Wino system is still well-separated in mass from the Higgsino system, decays between two systems still produce energetic $Z$ bosons. Thus, the Goldstone equivalence theorem again relates the partial widths of the heavier system's decay. 

The mass eigenstates of the Bino-Wino system are photinos and Zinos -- they are indistinguishable. Photinos do not couple to $Z$ nor $h$ as photons do not (at tree-level). Thus the situation is similar to previous cases as if photinos were absent and Zinos were the only gauginos in the system. Indeed, the Zino's partial width ratio is just given by the same formula in \Eq{eq:ew-higgsino}. Decays of charginos also produce $Z$ and $h$, but their contributions are not different from those of the Wino-Higgsino case. Therefore, the same observable relations in \Eq{eq:conc-lsp} and \Eq{eq:conc-nlsp} follow; $R_{Z/h} \simeq 1$.

We now extend the discussion by considering non-MSSM neutralinos: DFSZ axinos~\cite{Dine:1981rt} and gravitinos. They are weakly interacting, so they are relevant at collider only if they are LSPs. But in our formal discussion here, we do not stick to LSP cases. They have slightly different coupling structures to Higgsino sectors than those of electroweakinos, so we can enlighten how our conclusion is still drawn independently on these structures.

For the axino-Higgsino case, we have
\beq
\frac{\Gamma(\chi_i^0 \to \chi_j^0 Z)}{\Gamma(\chi_i^0 \to \chi_j^0 h)} \, \simeq \, \frac{ |s_\beta N_{H_k 3} - c_\beta N_{H_k 4} |^2 \, ( 1-2 \sqrt{r_j} ) }{ | s_\beta N_{H_k 3} + c_\beta N_{H_k 4} |^2 \, ( 1+2 \sqrt{r_j} ) }.
\label{eq:axi-higgsino} \eeq
This is obtained from \Eq{eq:neu-gold} by using couplings in \Eq{eq:nnphi}. The Goldstone equivalence theorem and the \Eq{eq:neu-gold} are not modified by the existence of DFSZ axinos as discussed in Appendix \ref{sec:app-gold}. By using \Eq{eq:higgsinomixing}, we again obtain the same observable relation in \Eq{eq:conc-lsp} and \Eq{eq:conc-nlsp}; $R_{Z/h} \simeq 1$.

One might expect that dominant axino-Higssino-$h$ couplings at tree-level would make Higgsinos dominantly decay into $h$. As discussed in \Sec{sec:prelim}, it is not always true. As energetic longitudinally-polarized $Z$ bosons are produced, the equivalence theorem applies and the Goldstone enhancement overcomes the small axino-Higgsino-$Z$ couplings.

For the gravitino-Higgsino case, we have
\beq
\frac{\Gamma(\chi_i^0 \to \chi_j^0 Z)}{\Gamma(\chi_i^0 \to \chi_j^0 h)} \, \simeq \, \frac{ |c_\beta N_{H_k 3} - s_\beta N_{H_k 4} |^2  }{ | c_\beta N_{H_k 3} + s_\beta N_{H_k 4} |^2 }.
\label{eq:grav-higgsino} \eeq
The gravitino results are present in, e.g., Refs.~\cite{Ambrosanio:1996jn,Dimopoulos:1996yq,Meade:2009qv}. By using \Eq{eq:higgsinomixing}, we again obtain the same observable relation in \Eq{eq:conc-lsp} and \Eq{eq:conc-nlsp}; $R_{Z/h} \simeq 1$.

One can note different relative signs and $t_\beta$ dependences among \Eq{eq:ew-higgsino}, \Eq{eq:axi-higgsino} and \Eq{eq:grav-higgsino}. The differences are inherited from different Higgsino couplings to electroweakinos, axinos and gravitinos. $\widetilde{H}_u$ and $\widetilde{H}_d$ couple to Winos and Binos with opposite sign due to opposite charges while both Higgsinos couple to gravitinos with the same sign. This explains the relative sign difference between \Eq{eq:ew-higgsino} and \Eq{eq:grav-higgsino}. On the other hand, the axino couples to different types of Higgsinos, i.e. $\widetilde{H}_u H_d$ and $\widetilde{H}_d H_u$, while other inos couple to the same type of Higgsinos. This introduces different $t_\beta$ dependence for the axino case in \Eq{eq:axi-higgsino}. 

In spite of these differences, the mixing angles still satisfy the relation \Eq{eq:higgsinomixing} and  our conclusions expressed in \Eq{eq:conc-lsp} and \Eq{eq:conc-nlsp} apply to all these cases.

\subsection{In the absence of Higgsinos}

It has been discussed~\cite{Gunion:1987yh} that, in the absence of Higgsinos, decays between well-separated Binos and Winos are dominantly through $h$ rather than $Z$. This is often true even for other cases of LSPs and NLSPs but not always true. In this subsection, we derive and generalize the result again using the Goldstone equivalence theorem, and we will argue that $R_{Z/h} \simeq 1$ is not a generic prediction of the absence of Higgsinos. 

For the Bino-Wino case, the Goldstone equivalence theorem still plays an important role in determining the branching ratios(BR) of the NLSP because gauginos couple to $Z$ bosons via Higgsino mixtures. The width ratio is given by (specifically for the Wino-LSP case, for instance)
\beq
\frac{\Gamma(\chi_i^0 \to \chi_j^0 Z)}{\Gamma(\chi_i^0 \to \chi_j^0 h)} \, \simeq \, \frac{ | (s_\beta N_{i4} + c_\beta N_{i3}) - t_W ( s_\beta N_{j4} + c_\beta N_{j3}) |^2 \, ( 1-2 \sqrt{r_j}) }{ | (s_\beta N_{i4} - c_\beta N_{i3}) - t_W ( s_\beta N_{j4} - c_\beta N_{j3}) |^2 \, ( 1+ 2\sqrt{r_j})}.
\label{eq:ratnohiggs}  \eeq

To proceed further, we look at the detailed forms of the $N_{ij}$. Mixing angles are approximated in the split limit as \cite{Gunion:1987yh}
\beq
\left( \bmat \frac{ m_Z s_W( M_1 c_\beta + \mu s_\beta )}{ (\mu^2 - M_1^2 )} &  \frac{-m_Z s_W( M_1 s_\beta + \mu c_\beta )}{ (\mu^2 - M_1^2 )} \\  \frac{-m_Z c_W( M_2 c_\beta + \mu s_\beta )}{ (\mu^2 - M_2^2 )} & \frac{ m_Z c_W( M_2 s_\beta + \mu c_\beta )}{ (\mu^2 - M_2^2 )} \emat \right),
\label{eq:mixing-winobino}\eeq
where the rows correspond to Bino-like and Wino-like neturalino mass eigenstates, respectively, and the columns correspond to $\widetilde{H}_d^0$ and $\widetilde{H}_u^0$ interaction eigenstates, respectively. We assume that gaugino masses $M_{1,2}$ are positive while we allow $\mu$ can be both positive and negative\footnote{Only relative phases, arg($\mu M_i$), are physical.}. For a positive $\mu >0$, the signs of each elements are fixed independently on $t_\beta$ and the size of $\mu$ (as long as $\mu > M_{1,2}$): $N_{i4}/N_{i3} <0$ and $N_{i4}/N_{j4}<0$. These relative signs imply that, in \Eq{eq:ratnohiggs}, the coupling square factor in the denominator is always greater than that in the numerator. This explains why decays into $h$ is always dominant.

But, for a negative $\mu<0$, the signs of mixing angles are not fixed when ${\rm min}(M_1,M_2)/t_\beta \lesssim \mu \lesssim {\rm max}(M_1,M_2) \, t_\beta$. We will numerically analyze the behavior of \Eq{eq:ratnohiggs} with a negative $\mu$ in \Sec{sec:rzh}, and we will see that $R_{Z/h} \gg 1$ is possible when the accidental cancellation between various terms in \Eq{eq:ratnohiggs} occurs. But $R_{Z/h} \simeq 1$ is not a generic prediction of this case, anyway.

The axino-gaugino case is similar to the Bino-Wino case in the sense that axinos and gauginos couple only via Higgsino mixtures. In this case, however, the ratio is not approximated by a simple formula as a very small number $\sim \mu /  v_{PQ}$ is involved and even a small mixing angle can give some non-negligible contributions. Instead, we carry out full numerical study in \Sec{sec:rzh}. We will also see that $R_{Z/h} \simeq 1$ is not a generic prediction of this case.

The gravitino-gaugino case is different, but $R_{Z/h} \simeq 1$ is still not a prediction of the case. Gravitinos differ from neutralinos in the sense that they can couple to gauginos without Higgsino mixtures; gauginos couple to gravitinos with transverse gauge bosons while Higgsinos couple with the longitudinal components, so they do not interfere~\cite{Dimopoulos:1996yq}. As a result, pure gaugino's decay into gravitinos at high-energy is not approximated by the equivalence theorem. The BRs into $Z$ and $\gamma$ are rather fixed by the weak mixing angle $c_W$ as~\cite{Ambrosanio:1996jn,Dimopoulos:1996yq,Meade:2009qv}
\beq
\frac{ \Gamma( \chi_1^0 \to \gamma \widetilde{G} ) }{ \Gamma(\chi_1^0 \to Z \widetilde{G}) } \= \frac{ | c_W N_{11} + s_W N_{12} |^2}{| s_W N_{11} - c_W N_{12} |^2}.
\eeq
Thus, $s_W^2 \sim 23\%$ of a Wino-NLSP decays to $\gamma$ and the remaining decays to $Z$. A Bino-NLSP has the opposite BRs. If we observe such specific BRs, it would be a useful indication of the gravitino-gaugino case and the Higgsinos' absence.

%%%%%%
\section{NLSP Higgsino productions} \label{sec:production}

It is necessary that two neutral Higgsino NLSPs (if Higgsinos are NLSPs) should be produced by equal numbers. Otherwise, total NLSP decays would produce different numbers of $h$ and $Z$.

There are three production channels of NLSP Higgsino pairs in the split limit: (1) $pp \to Z^* \to \widetilde{H}_i^0 \widetilde{H}_j^0$, (2) $pp \to Z^*/ \gamma^* \to \widetilde{H}^+ \widetilde{H}^-$ and (3) $pp \to W^* \to \widetilde{H}_i^0 \widetilde{H}^\pm$. For the process (1), the process is non-vanishing only for $i\ne j$, so always two different neutral Higgsinos are produced together, hence the same number. This can be easily seen from that $\sigma(pp \to \widetilde{H}_i^0 \widetilde{H}_j^0) \, \propto \, ({\cal O}^{\prime \prime L}_{ij})^2$ (if no CP-phases) and that the couplings become
\beq
{\cal O}^{\prime \prime L}_{ii} \, \propto \, N_{i3} N_{i3} - N_{i4}N_{i4} \, \propto \, \frac{1}{2} - \frac{1}{2} \=0
\ceq
{\cal O}^{\prime \prime L}_{ij} \, \propto \, N_{i3} N_{j3} - N_{i4}N_{j4} \, \propto \, \frac{1}{2} + \frac{1}{2} \=1
\eeq
where we have used the relation \Eq{eq:higgsinomixing} in the second equation. For the process (2), charginos are not a problem because it decays equally to $h$ and $Z$. The process (3) is equal for $i=1$ and $2$, hence again the same number of two neutral Higgsinos. To see this, we first note that the production rate contains three pieces
\beq
\sigma( pp \to \widetilde{H}_i^0 \widetilde{H}_j^+) \quad \propto \quad  m_{\chi_i^0} Re[ {\cal O}^L_{ij} {\cal O}^{R *}_{ij} ], \quad | {\cal O}^L_{ij} |^2, \quad | {\cal O}^R_{ij} |^2.
\eeq
Each piece is the same for $i=1$ and 2 because
\beq
\frac{{\cal O}^{L}_{ij} }{ {\cal O}^{R *}_{ij} } \, \simeq\, - \frac{N_{i4}}{ N_{i3}}, \quad \textrm{and} \quad \frac{ m_{\widetilde{H}_1^0} }{ m_{\widetilde{H}_2^0} } \, \simeq \, -1
\eeq
and because of the relation \Eq{eq:higgsinomixing}. The opposite sign of the Higgsino mass eigenvalues can also be easily understood from the $2 \times 2$ sub-matrix in \Eq{eq:higgsino2by2}.

In all, the same numbers of two neutral Higgsinos are directly produced. This completes our formal derivation of the observable signal at LHC $pp$ collider. The discussion is also valid at $e^+ e^-$ colliders although the measurability of $R_{Z/h} \simeq 1$ is a different question.

NLSP-gaugino pair productions are not concerning. There will only be one NLSP neutralino, and its decays into two neutral Higgsinos automatically satisfy the desired relations \Eq{eq:conc-lsp} and \Eq{eq:conc-nlsp}. Even for the nearly degenerate Bino-Wino case, weakly interacting photinos are not produced much and essentially the decay of Zinos generate the signal.

%%%%%%
\section{$R_{Z/h}$} \label{sec:rzh}

We define the collider observable 
\beq
R_{Z/h} \, \equiv \, \frac{ \sum_{i,j  } \sigma(\chi_i) \times {\rm BR}(\chi_i \to \chi_j + Z) }{ \sum_{i,j} \sigma(\chi_i) \times {\rm BR}(\chi_i \to \chi_j + h) }
\eeq
which, if a luminosity is multiplied, really counts (and takes the ratio) the numbers of $Z$ and $h$ bosons produced from all possible indistinguishable NLSP productions and subsequent decays. The notations $i \in \{NLSP\}$ and $j \in \{LSP\}$ sum over all indistinguishable NLSP $i$ and all indistinguishable LSP $j$. In the numerical study, we conveniently define all ino states within 20GeV-mass gap to be indistinguishable inos. A factor 2 has to be properly multiplied if the same particles are pair produced such as in $\chi_i^+ \chi_i^-$.

In this section, we numerically calculate the observable without any formal approximation at the LHC14 and demonstrate the formal discussions in previous sections. 

We are envisaging ideal measurements. We do not carry out full collider simulation nor do we take into account backgrounds. But we briefly discuss measurement prospects at the end of this section.

\subsection{Electroweakinos}

\begin{figure}[t] \centering
\includegraphics[width=0.49\textwidth]{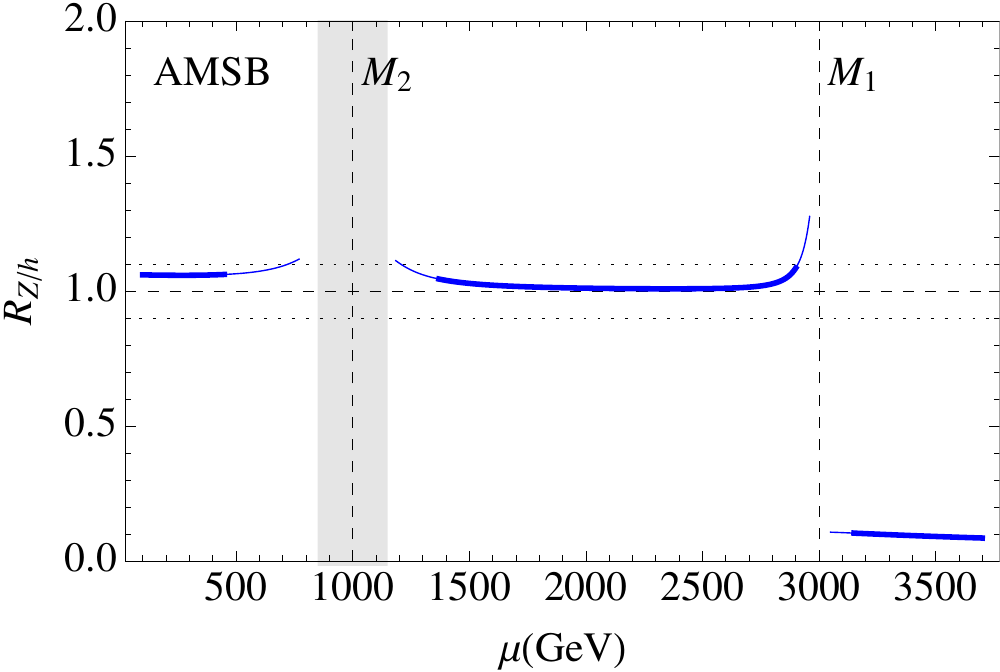}
\includegraphics[width=0.49\textwidth]{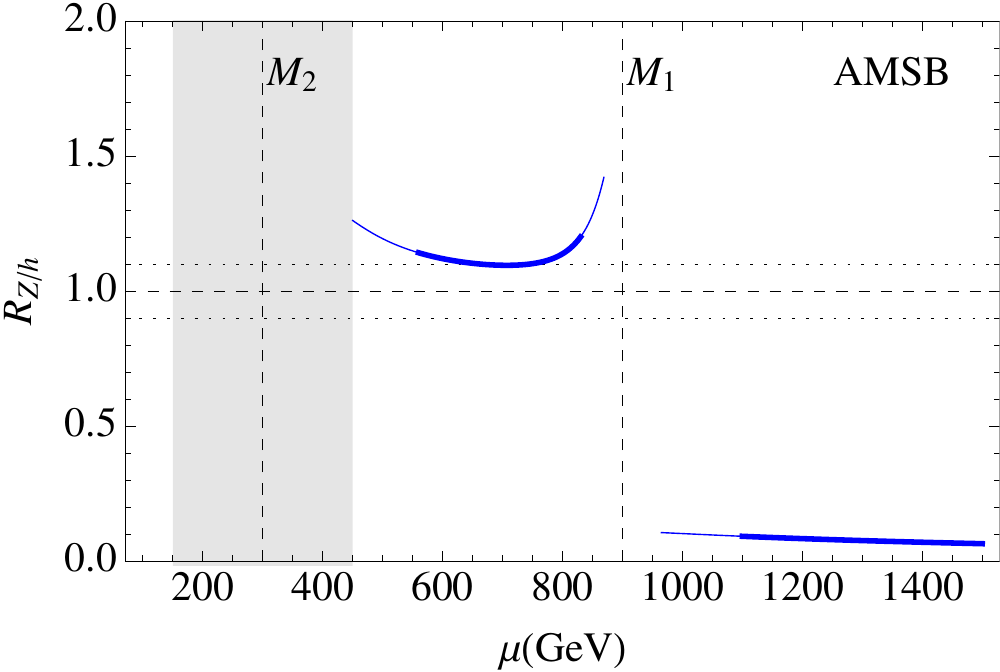}
\caption{$R_{Z/h}$ in AMSB. The gaugino mass ratio, $M_2:M_1 =1:3$, is assumed. $M_2=$1000GeV(left) and 300GeV(right). The thick(thin) blue line is $R_{Z/h}$ for $\Delta m <$10(20) GeV among two neutral Higgsinos. $t_\beta=5$ and $\mu>0$. Gray regions are not considered because two-body decays of NLSPs into $h$ are not allowed. Horizontal dashed lines are for reference at $R_{Z/h}=1.0\pm0.1$.  Vertical dashed lines are at $M_{1,2}$. When the Higgsino is the LSP or NLSP, $R_{Z/h}$ is close to 1 while heavy Higgsinos imply dominant NLSP decays into $h$. }
\label{fig:amsb}
\end{figure}

We study three scenarios of the electroweakino-Higgsino case depending on the relative mass orderings of electroweakinos. In the Anomaly-mediated(AMSB)\cite{Randall:1998uk} scenario, the Wino is the lightest and the Bino is about three times heavier. In the minimal supergravity(mSUGRA) scenario, the Bino is the lightest and the Wino is about twice heavier; the mass hierarchy is smaller than the AMSB case. We also consider the compressed spectrum where the Wino and the Bino are nearly degenerate forming an indistinguishable set of states, i.e. $M_1 \simeq M_2$. In all cases, we assume that gluinos are heavier.

\begin{figure}[t] \centering
\includegraphics[width=0.49\textwidth]{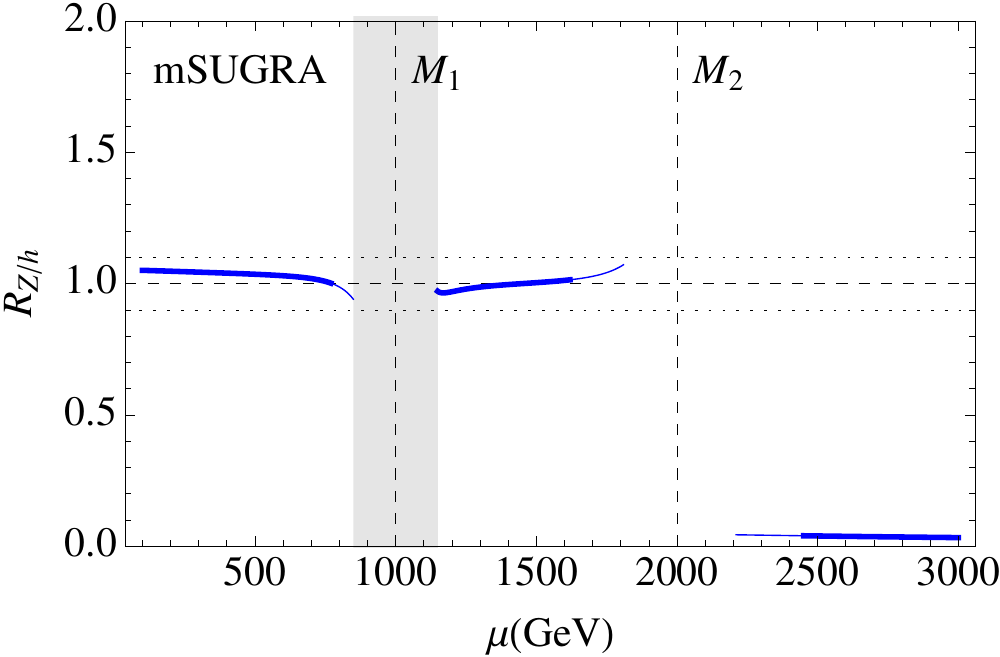}
\includegraphics[width=0.49\textwidth]{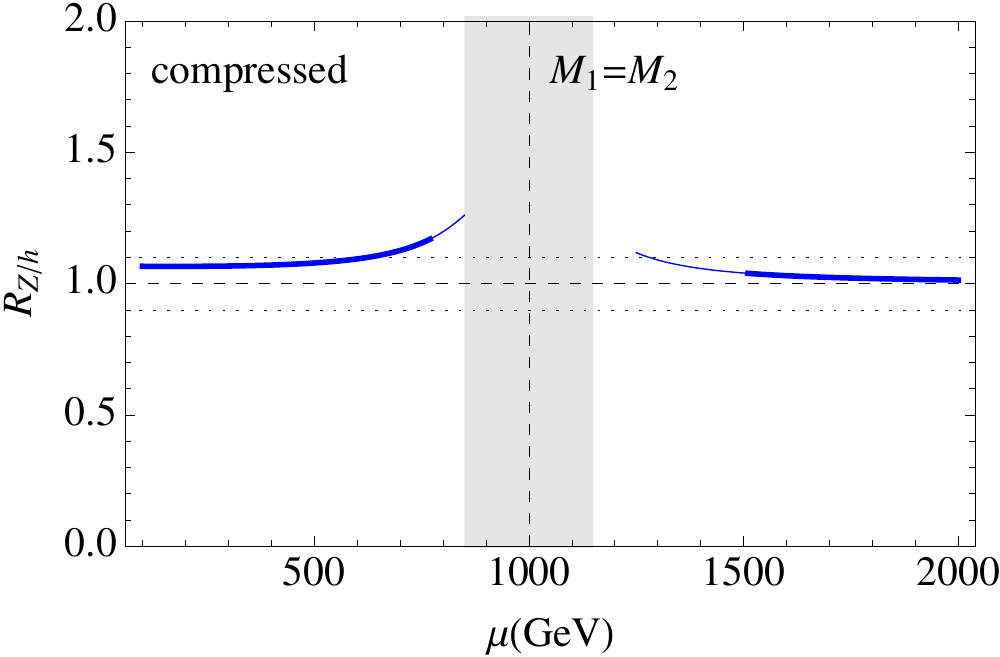}
\caption{$R_{Z/h}$ in mSUGRA(left) and compressed spectrum(right). The mSUGRA gaugino mass ratio, $M_1:M_2 =1:2$, is assumed with $M_1=1000$ GeV in the left panel. In the right panel, we consider a compressed spectrum with $M_1=M_2=1000$ GeV. $t_\beta=5$ and $\mu>0$. All other plot details are as in \Fig{fig:amsb}. When the Higgsino is the LSP or NLSP, $R_{Z/h}$ is close to 1 while heavy Higgsinos imply dominant NLSP decays into $h$.}
\label{fig:msugra}
\end{figure}

The observable $R_{Z/h}$ is plotted for each three scenarios in \Fig{fig:amsb} and \Fig{fig:msugra}. When the Higgsino is the LSP or the NLSP, the advocated signal $R_{Z/h} \simeq 1$ is obtained in all scenarios. The signal is independent on the ordering of gaugino masses. When the gauginos are relatively light and the mass-gap is correspondingly small (compare two panels in \Fig{fig:amsb}), the Goldstone equivalence theorem is less accurate and the deviation of $R_{Z/h}$ from the unity becomes larger. But it is still close to the unity and can be distinguishable from heavy Higgsino cases. We mention again that, for the compressed spectrum in \Fig{fig:msugra}, we add all indistinguishable productions and decays between the Higgsino system and the Bino-Wino system, and $R_{Z/h} \simeq 1$ is obtained.

We also see that, as long as two-body decays of the NLSP to on-shell $h$ bosons are allowed, the mass splittings among Higgsino states are small, $\Delta \lesssim 20$GeV, and all Higgsino states are virtually indistinguishable. The signal is then valid and the equivalence theorem is a good approximation.

\begin{figure}[t] \centering
\includegraphics[width=0.49\textwidth]{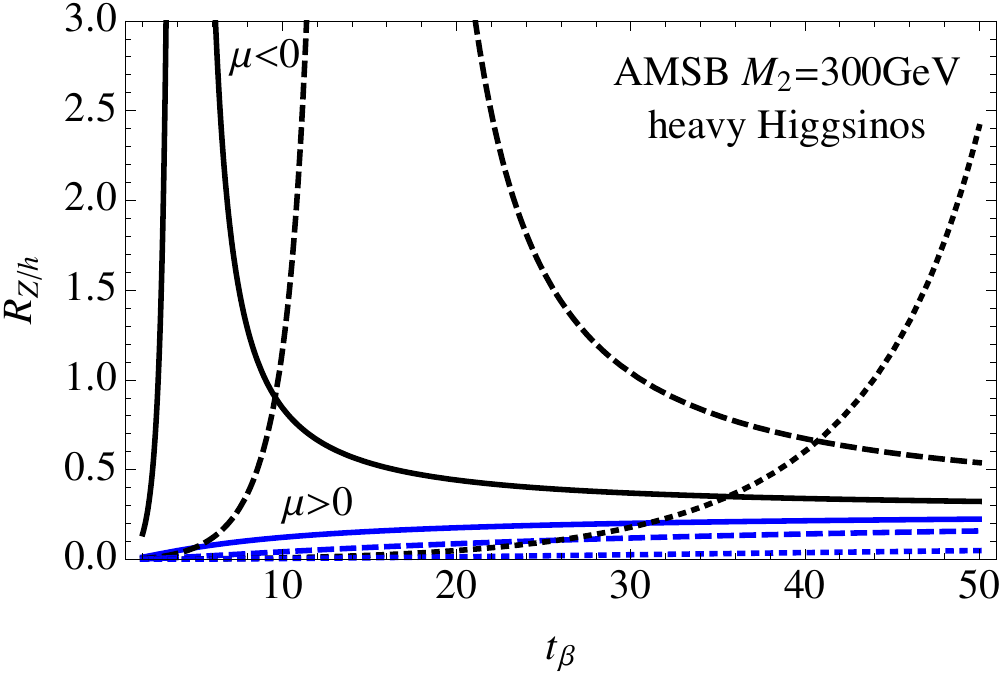}
\includegraphics[width=0.49\textwidth]{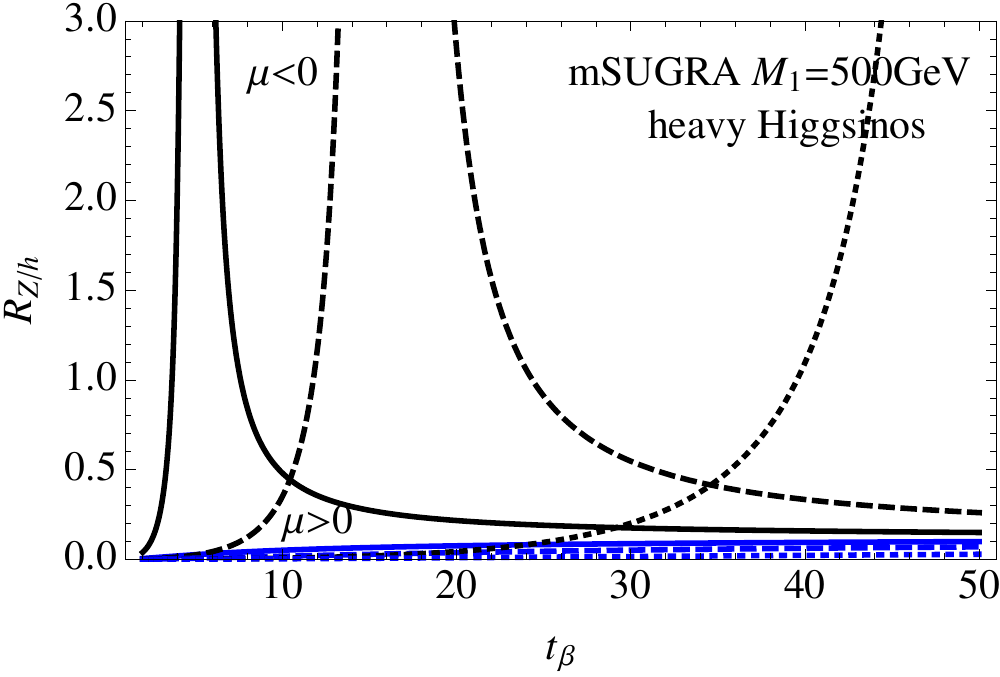}
\caption{$R_{Z/h}$ for heavy Higgsinos with positive(blue) and negative(black) $\mu$. In the left panel, the AMSB is asusmed with $M_2=300$GeV and $|\mu|=$1.5, 4.5, 20TeV (from left to right lines). In the right panel, mSUGRA is assumed with $M_1=500$GeV and $|\mu|=$2, 6, 20TeV. $R_{Z/h}$ is always small for positive $\mu$. On the other hand, $R_{Z/h}$ can be much larger than 1 for some regions of negative $\mu$ and $t_\beta$ due to the accidental cancellation discussed in text. In any case, $R_{Z/h} \simeq 1$ is not a generic prediction of the heavy Higgsinos.}
\label{fig:heavyh}
\end{figure}

When the Higgsino is heavy (NNLSP or higher), $R_{Z/h} \ll 1$ is obtained in all cases with $\mu >0$; see \Fig{fig:amsb} and \Fig{fig:msugra} and also \Fig{fig:heavyh}. Such small $R_{Z/h}$ will be easily distinguishable from $R_{Z/h} \simeq 1$. The result is also independent on the relative mass ordering of gauginos.

It is also possible to have a negative $\mu$. The sign of $\mu$ does not change the conclusion of obtaining $R_{Z/h} \simeq 1$ for light Higgsinos. We, however, expected that the heavy Higgsino (NNLSP or higher) case is sensitive to the sign of $\mu$. \Fig{fig:heavyh} demonstrates that, for an intermediate negative $\mu<0$, the $R_{Z/h}$ can be sizably larger than 1. $R_{Z/h}$ becomes small again for very heavy Higgsinos or small or large $t_\beta$. The behavior (compare two panels in \Fig{fig:heavyh}) is also independent on the ordering of gaugino masses. In any case, $R_{Z/h} \simeq 1$ is not a generic prediction of the heavy Higgsino with a negative $\mu$.    

The behavior of $R_{Z/h}$ in \Fig{fig:heavyh} can be understood from \Eq{eq:ratnohiggs} and \Eq{eq:mixing-winobino}. The relative signs of mixing angles in \Eq{eq:mixing-winobino} largely determine whether the ratio \Eq{eq:ratnohiggs} is larger than or smaller than 1. When $t_\beta \to 1$ or negative $|\mu| \to {\rm max}(M_1,M_2) t_\beta$ becomes large enough, signs of all four mixing angles flip compared to those with positive $\mu$, hence small $R_{Z/h}$ as with positive $\mu$. When $t_\beta \to \infty$, only $N_{i4}$ and $N_{j4}$ terms are important in \Eq{eq:ratnohiggs} while they have opposite signs $N_{i4}/N_{j4}<0$. Thus, those terms add up to a larger value in the denominator. These explain the sharping behavior of $R_{Z/h}$. 

We finally comment that the signal may not be observable in the Bino NLSP case due to Bino's small productions. Then the NNLSP production and decay can be important. As long as the NNLSP is abundantly produced and dominantly decays directly to the LSP, our results will apply as if the NNLSP were the NLSP -- but in this case, we may miss the existence of Binos.

In all cases, if we observe $R_{Z/h} \simeq 1$, it strongly indicates the existence of Higgsinos as the LSPs or NLSPs.

\subsection{Competing decay modes and weakly interacting LSPs}

Our argument relies on the fact that the heavier neutral Higgsino does not dominantly decay into the lighter neutral Higgsino (when Higgsinos are NLSPs). In the split limit, the mass splitting between two neutral Higgsinos, $\Delta$, is much smaller than $m_{Z}$, so the decays between Higgsinos are suppressed by small three-body phase space. However, when the LSP is a weakly interacting particle, this phase space suppression may not be enough.

\begin{figure}[t] \centering
\includegraphics[width=0.49\textwidth]{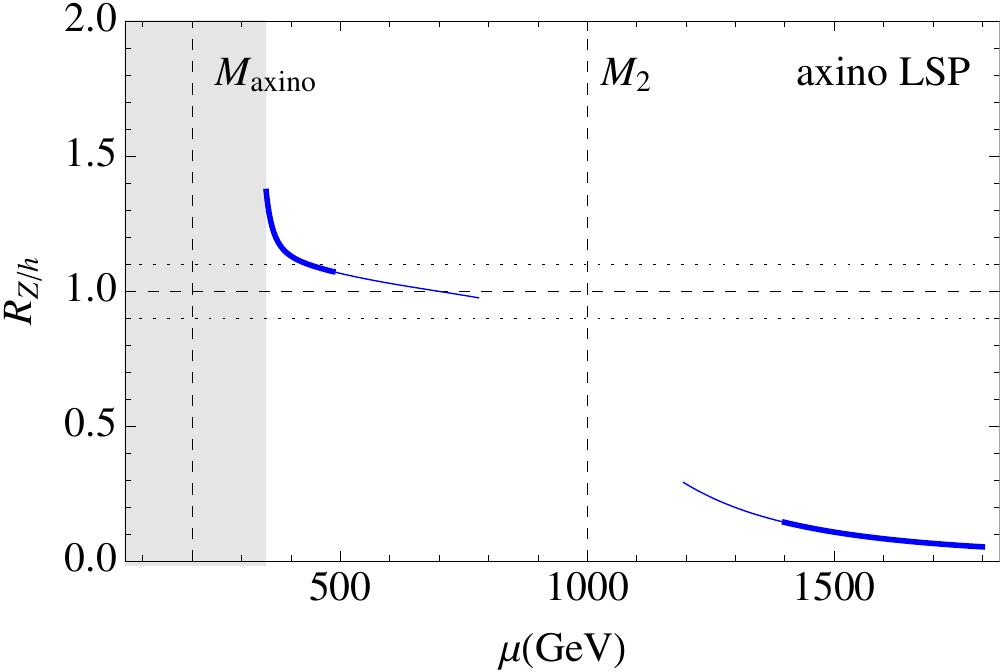}
\includegraphics[width=0.49\textwidth]{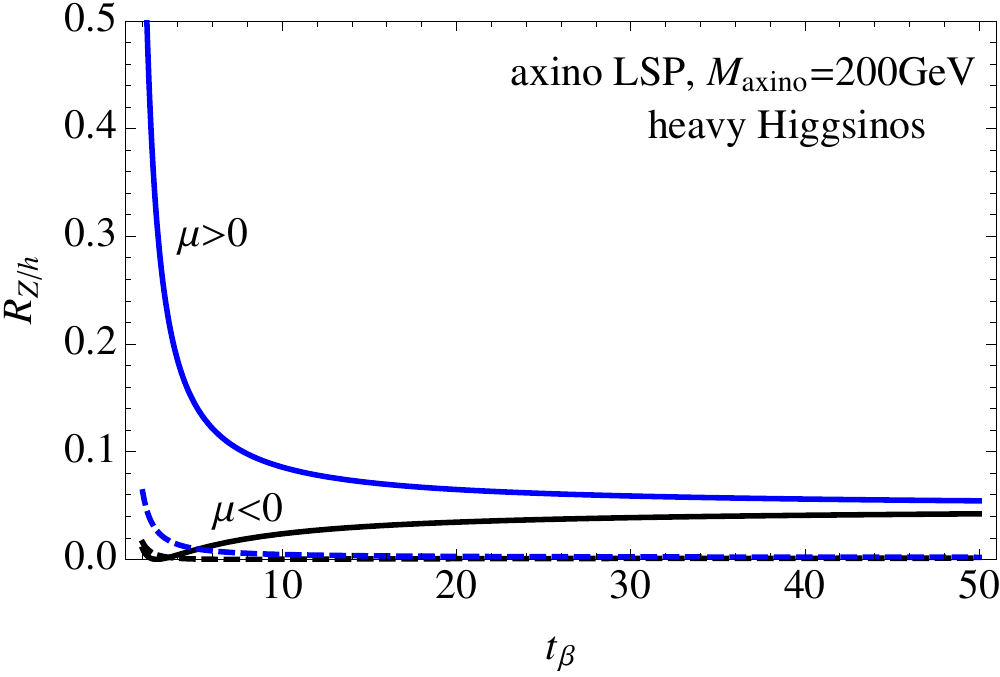}
\caption{$R_{Z/h}$ in the case of axino LSP. The axino mass is 200GeV and $M_2=1$TeV in both panels. Binos are heavy and decoupled for simplicity. All other details of the left panel are same as in \Fig{fig:amsb}. In the right panel, $|\mu|=1.4, 3$ TeV from top to bottom lines. All other details of the right panel are same as in \Fig{fig:heavyh}.}
\label{fig:axino}
\end{figure}

Taking the DFSZ axino as a weakly interacting example, we checked that the Winos and Binos should be somewhat heavier than Higgsinos so that the mass splitting, $\Delta$, between Higgsinos is small enough to sufficiently suppress the three-body decay width which scales with $\Delta^5$. Which decay mode dominates actually depends on various model parameters and even on other particle masses if one considers the loop-induced two-body decay into photons. Due to these model dependencies and given the possibility for the desired decay modes to dominate, we will not further consider this issue; we assume that all decays between NLSPs and LSPs are prompt and dominant. We refer to Appendix \ref{sec:app-int} for more discussions, and a related detailed study will be presented in our future publication~\cite{SHJung}. 

$R_{Z/h}$ for the axino LSP is calculated in \Fig{fig:axino}. Again, the Higgsino NLSP predicts $R_{Z/h} \simeq 1$ while heavy Higgsinos predict small $R_{Z/h} \ll 1$. Unlike previous cases, a negative $\mu$ with heavy Higgsinos also predicts small $R_{Z/h}$. For the gravitino-Higgsino case, we also obtain $R_{Z/h} \simeq 1$ for light Higgsinos although we do not present numerical results for this case.

These results also imply that it will not be easy to know whether the LSP is axinos or gravitinos or other neutralinos if Higgsinos are NLSPs. This may add new degeneracies to the LHC inverse problem.

%%%
\subsection{Measurement prospects}

Finally, we briefly discuss which part of parameter space depicted in Fig.~\ref{fig:amsb},\ref{fig:msugra} and \ref{fig:axino} is practically relevant to future LHC experiments.

The discovery of NLSP inos will first likely be made by utilizing chargino-neutralino NLSP pair production with their subsequent decays to the LSP via $WZ$ or $Wh$~\cite{Aad:2014nua, TheATLAScollaboration:2013zia,Aad:2014vma,CMS:2013afa,CMS:2013dea,Khachatryan:2014qwa}. We conveniently use these two channels for the discussion in this subsection. The $5\sigma$ discovery prospect has been estimated in Ref.~\cite{Han:2013kza}. For the case of Bino-LSP and either Higgsino- or Wino-NLSP, for example, the discovery is expected to be possible up to 350GeV at LHC14 with 300fb$^{-1}$ by taking into account proper branching ratios of NLSPs and combining $WZ$ and $Wh$ channels. By naively scaling this discovery reach\footnote{The scaling of the reach from one measurement to another (different collision energy and luminosity applied to different parameter space of new physics) is approximately possible especially in the split limit because searches rely on the high-energy regions of $M_{eff}$ or MET or similar kinds, and NLSPs are essentially massless in this phase space~\cite{Jung:2013zya}. We use \texttt{Collider Reach ($\beta$)} tool for numerical scaling~\cite{Salam:reach-scaling}.}, we obtain a discovery reach extended to 1200GeV at LHC100 with 1ab$^{-1}$. Thus any parameter space with NLSPs heavier than about 1200GeV, or equivalently $\mu \gtrsim 1200$GeV in Fig.~\ref{fig:amsb},\ref{fig:msugra} and \ref{fig:axino}, may not be that practically relevant. 

We then compare the measurement prospects of $Z$ and $h$ bosons from the results of $WZ$ and $Wh$ channels. The $WZ$ channel is expected to be better; the current LHC exclusion limit from the $WZ$ channel is stronger than that from the $Wh$ channel~\cite{Aad:2014nua, TheATLAScollaboration:2013zia,Aad:2014vma,CMS:2013afa,CMS:2013dea,Khachatryan:2014qwa}.  Thus, $Z$ bosons will be first and better measured than $h$ bosons (if branching ratios are similar). 

After the discovery in the $Wh$ channel, one will then be able to measure $R_{Z/h}$. The distinction of $WZ$ and $Wh$ channels, hence the distinction of $Z$ and $h$ allowing the measurement of $R_{Z/h}$, will be possible when both channels are efficiently measured with three leptons plus MET. The resonance peak of the $Z$ boson is formed from same-flavor-opposite-sign lepton pairs (among three leptons) while the $h$ resonance is not formed due to neutrinos from Higgs decays. If the $WZ$ channel is not discovered by the discovery of $Wh$, it would already strongly suggest that the Higgs boson is much more produced than the $Z$ boson, i.e., $R_{Z/h} \ll 1$.

Future linear collider can be another place to efficiently measure the $R_{Z/h}$ signal. Although the discovery reach would not extend much beyond that of LHC100, a higher precision of the measurement maybe obtained.

%%%%%%%
\section{Conclusions and discussions}

We have studied the distinctive signature of the NLSP decay that can be a confident evidence of the existence of light Higgsinos in the split limit.  When Higgsinos are the LSPs or NLPSs, the equal numbers of $Z$ bosons and $h$ bosons will always be present in the NLSP production and decay; $R_{Z/h} \simeq 1$. On the other hand, heavier Higgsinos (NNLSPs or higher) typically predict the ratio, $R_{Z/h}$, much smaller than or larger than 1.  As an illustration, two models introduced as degenerate models in the introduction, Table~\ref{tab:inverse}, now predict very different $R_{Z/h}=1.03, 5.35$ for Higgsino- and Wino-NLSP models; they will be easily distinguishable. 

The BRs of a single neutralino, of course, depend on the identity of the neutralino. But they also depend on other parameters of a theory such as neturalino mixing angles and $t_\beta$. Thus, a mere BR is not an efficient tagger of the identity. In the split limit, however, two neutral Higgsinos (and a charged one) are nearly degenerate and indistinguishable at collider. Thus, all indistinguishable production and decay processes involving Higgsinos add up to generate the observable signal $R_{Z/h} \simeq 1$ independently from other parameters of a theory. We analytically derived the observable using the Goldstone equivalence theorem and numerically demonstrated the relation without any approximations. 

%How well can one measure the Higgs boson final states compared to the $Z$ boson final states? The current constraints on electroweakino parameter space coming from the Higgs boson final states are weaker than those from $W,Z$ final states~\cite{Aad:2014nua, TheATLAScollaboration:2013zia,Aad:2014vma,CMS:2013afa,CMS:2013dea,Khachatryan:2014qwa}. According to a LHC14 prospect study~\cite{Han:2013kza}, a $5\sigma$ discovery is expected up to $\mu \sim 200$GeV with 300fb$^{-1}$ in the split limit. This discovery reach is again lower than that from the $W,Z$ final states, $400\sim 500$GeV~\cite{ATLAS:2013hta}. It will thus be useful to improve the search in the Higgs boson final states or to invent a discriminating observable in the $W,Z$ sector alone such as $R_{W/Z}$. Future linear collider can also be a place to efficiently measure the $R_{Z/h}$ signal; see Ref.\cite{Berger:2007yu} for other strategies at linear collider.

Exceptions to the discussion, however, may arise when the LSP is weakly interacting such as axinos or gravitinos and the heavier neutral Higgsino dominantly decays to the lighter neutral Higgsino. When this does not happen, the same signal $R_{Z/h} \simeq 1$ is observable for these cases too. Although the existence of Higgsinos can be established in this way, the existence of such non-MSSM neutralinos may add other degeneracies to the inverse problem.

Our method still leaves the two-fold degeneracy in the ino spectrum; Higgsinos can be either the LSPs or NLSPs. So once the existence of Higgsinos is estbailshed, a dedicated $\chi^2$ analysis or other strategies will further be needed to lift the degeneracy. 

All our discussions are made in the split limit. It is perhaps the most difficult scenario for the discovery. It could still well be that inos sizably mix and there are many particles separated only by reasonable mass gaps in the accessible spectrum. Then the discovery and precision measurements will be easier although there could still be some residual inverse problems~\cite{Balazs:2009it}.  In any case, we believe that our formal discussions based on the Goldstone equivalence theorem will be useful in obtaining insights on various possible decay modes and in resolving the (residual) inverse problem.

%%%%%%%
\vspace{0.2cm}
{\bf \emph{Acknowledgement.}} We acknowledge E.J.Chun, H.M.Lee, W.I.Park, C.S.Shin, L.T.Wang for many useful discussions. We thank E.J.Chun and J.Wells for reading the manuscript carefully. We also thank L.T.Wang for reminding us of the LHC inverse problem. The work is supported in part by National Research Foundation (NRF) of Korea under grant 2013R1A1A2058449.

%%%%%%%
\appendix
\section{Interactions and decay widths} \label{sec:app-int}

Here we collect analytic expressions used throughout this paper.

The interaction Lagrangian of inos in terms of mass eigenstates is
\bea
{\cal L} &=& g \overline{\chi_i^0} \gamma^\mu \left( {\cal O}_{ij}^L P_L + {\cal O}_{ij}^R P_R \right) \chi_j^+ W_\mu^+ + h.c. \nonumber\\ && +\frac{g}{c_W} \overline{\chi_i^+} \gamma^\mu \left( {\cal O}_{ij}^{\prime L} P_L + {\cal O}_{ij}^{\prime R} P_R \right) \chi_j^+ Z_\mu + \frac{g}{c_W} \overline{\chi^0_i} \gamma^\mu ( {\cal O}^{\prime \prime L}_{ij} P_L + {\cal O}^{\prime \prime R}_{ij} P_R ) \chi^0_j Z_\mu \nonumber\\
&& + g \, \overline{\chi_i^+} \left( D_{hij}^L P_L + D_{hij}^R P_R \right) \chi_j^+ h \,+ g \overline{\chi_i^0} \left( D^{\prime L}_{hij} P_L + D^{\prime R}_{hij} P_R \right) \chi_j^0 h
\label{lagran} \eea
where $h$ is the 125GeV Higgs boson. Couplings are give by
\bea
{\cal O}^L_{ij} &=& N_{i2} V^*_{j1} - \frac{1}{\sqrt{2}}N_{i4}V^*_{j2} , \qquad {\cal O}^R_{ij} = N^*_{i2} U_{j1} + \frac{1}{\sqrt{2}}N^*_{i3}U_{j2}  \nonumber\\
{\cal O}^{\prime L}_{ij} &=& -\delta_{ij} c_W^2 + \frac{1}{2} V_{i2}V^*_{j2}, \qquad {\cal O}^{\prime R}_{ij} = -\delta_{ij} c_W^2 + \frac{1}{2} U^*_{i2}U_{j2} \nonumber\\
{\cal O}^{\prime \prime L}_{ij} &=& -\frac{1}{2} ( N^*_{i3} N_{j3} - N^*_{i4} N_{j4} ) ,\qquad {\cal O}^{\prime \prime R}_{ij} = -({\cal O}^{\prime \prime L}_{ij})^* = -{\cal O}^{\prime \prime L}_{ji} \nonumber\\
D^L_{hij} &=& \frac{1}{\sqrt{2}}\left( -s_\beta U^*_{i1} V^*_{j2} - c_\beta U^*_{i2} V^*_{j1} \right) , \qquad D^R_{hij}  = (D^L_{hji})^* \nonumber\\
D^{\prime L}_{hij} &=& \frac{1}{2}\left( N^*_{j2}- t_WN^*_{j1} \right) \, \left(  N^*_{i4} s_\beta - N^*_{i3} c_\beta \right) + \Delta D^\prime_{hij} + (i \leftrightarrow j) , \quad D^{\prime R}_{hij} = (D^{\prime L}_{hji} )^* = (D^{\prime L}_{hij} )^*  \nonumber\\
\Delta D^\prime_{hij} &=& \frac{c_H \mu }{\sqrt{2} g v_{PQ}} N_{i5} \left( - N_{j3} s_\beta - N_{j4} c_\beta  \right).
\label{couplings} \eea
Here, we use the following basis of neutralinos 
\beq
\{ \, \widetilde{B}, \widetilde{W}^0, \widetilde{H}_d^0, \widetilde{H}_u^0, \axi \, \}.
\eeq
For Higgs couplings, we have already assumed a decoupling limit relevant to split SUSY: $c_\alpha(s_\alpha) \to -s_\beta(c_\beta)$. %For general Higgs couplings involving other Higgs bosons, refer to Appendix on Goldstone Equivalence theorem. 
$c_H=0$ recovers the MSSM results with only gauginos and higgsinos. Our notation without axinos conforms with that of Ref.~\cite{Martin:1997ns}. Details of DFSZ axino interactions can be found in, e.g., Ref.~\cite{Bae:2011iw}.

The two-body decay partial widths of an ino into other inos via $Z$ and $h$ are \cite{Gunion:1987yh,Martin:2000eq}
\bea
\Gamma(\chi_i^+ \to \chi_j^+ Z ) &=& \frac{g^2 \,  m_{\chi_i^+}}{32\pi c_W^2} \lambda^{1/2}(1,r_j, r_Z)   \cdot \Big[ \, \left( \left| {\cal O}^{\prime L}_{ij} \right|^2 + \left| {\cal O}^{\prime R}_{ij} \right|^2 \right) \left( \frac{(1-r_j)^2}{r_Z} + (1+r_j - 2r_Z)\right)  \nonumber\\
&& \qquad \qquad \qquad \qquad  - 12 Re[{\cal O}^{\prime L}_{ij} {\cal O}^{\prime R *}_{ij}] \sqrt{r_j} \, \Big]  \label{eq:dec-ch-Z}
\\
\Gamma(\chi_i^0 \to \chi_j^0 Z ) &=& \frac{g^2 \,  m_{\chi_i^0}}{16\pi c_W^2} \lambda^{1/2}(1,r_j, r_Z)   \nonumber\\
&& \qquad \cdot  \Big[ \, \left| {\cal O}^{\prime \prime L}_{ij} \right|^2 \left( \frac{(1-r_j)^2}{r_Z} + (1+r_j - 2r_Z) \right) + 6 Re[ ({\cal O}^{\prime \prime L}_{ij})^2]  \sqrt{r_j} \, \Big] \label{eq:dec-neu-Z}
\\
\Gamma(\chi_i^+ \to \chi_j^+ h ) &=& \frac{g^2 \,  m_{\chi_i^+}}{32\pi} \lambda^{1/2}(1,r_j, r_h) \cdot \Big[ \, \left( \left| D^{L}_{hij} \right|^2 + \left| D^{L}_{hji} \right|^2 \right) \left( 1+r_j-r_h \right) \nonumber\\
&& \qquad \qquad \qquad \qquad  + 4 Re[ D^{L}_{hij} D^{L}_{hji}] \sqrt{r_j} \, \Big] \label{eq:dec-ch-h}
\\
\Gamma(\chi_i^0 \to \chi_j^0 h ) &=& \frac{g^2 \,  m_{\chi_i^0}}{16\pi} \lambda^{1/2}(1,r_j, r_h) \cdot  \left( \left| D^{\prime L}_{hij} \right|^2  (1+r_j-r_h)  +  2 Re[ (D^{\prime L}_{hij})^2]\sqrt{r_j} \right) \label{eq:dec-neu-h}
\eea
where Re terms in scalar modes produce an extra minus sign for $A^0,G^0$. The two-body decay partial widths via $W$ bosons are \cite{Gunion:1987yh}
\bea
\Gamma(\chi_i^+ \to \chi_j^0 W^+ ) &=& \frac{g^2 \,  m_{\chi_i^+}}{32\pi } \lambda^{1/2}(1,r_j, r_W)  \cdot \Big[ \, \left( \left| {\cal O}^{L}_{ji} \right|^2 + \left| {\cal O}^{R}_{ji} \right|^2 \right) \left( \frac{(1-r_j)^2}{r_W} + (1+r_j - 2r_W)\right)  \nonumber\\
&& \qquad \qquad \qquad \qquad  - 12 Re[{\cal O}^{L}_{ji} {\cal O}^{R *}_{ji}] \sqrt{r_j} \, \Big] \label{eq:dec-ch-W}
\\
\Gamma(\chi_i^0 \to \chi_j^+ W^- ) &=& \frac{g^2 \,  m_{\chi_i^0}}{32\pi } \lambda^{1/2}(1,r_j, r_W)  \cdot \Big[ \, \left( \left| {\cal O}^{L}_{ij} \right|^2 + \left| {\cal O}^{R}_{ij} \right|^2 \right) \left( \frac{(1-r_j)^2}{r_W} + (1+r_j - 2r_W)\right)  \nonumber\\
&& \qquad \qquad \qquad \qquad  - 12 Re[{\cal O}^{L}_{ij} {\cal O}^{R *}_{ij}] \sqrt{r_j} \, \Big] \label{eq:dec-neu-W}
\eea
where $r_x \equiv m_x^2/m_{\chi_i}^2$ and $\lambda(x,y,z)=x^2+y^2+z^2-2xy-2yz-2zx$. Note $\sqrt{r_j}$ has the same sign as $m_j/m_i$, not always positive. No index summation of coupling factors.

The three-body decay widths via off-shell $W,Z$ gauge bosons are
\bea
\Gamma ( \chi_i^0 \to \chi_j^0 Z^* ) &=&  \frac{m_i}{192 \pi^3} \frac{g^4}{c_W^4} \, \int_0^{r_X^{max}} \, d r_X \, \lambda^{1/2}(1,r_X,r_j) \, \left( \sum_f (v_{Z,f}^2 + a_{Z,f}^2) N_c(f) \right) \nonumber\\
&&  \cdot \, \Big[ |{\cal O}^{\prime \prime L}_{ij}|^2  \left( ( 1-r_j)^2 + (1+r_j-2r_X)r_X \right) + 6 Re[ \left( {\cal O}^{\prime \prime L}_{ij} \right)^2 ] \, r_X \sqrt{r_j} \Big] \nonumber\\
&& \cdot \, \frac{1}{\left( (r_X-r_Z)^2 + r_Z r_{\Gamma_Z} \right)},
\eea
\bea
\Gamma ( \chi_i^+ \to \chi_j^0 W^* ) &=&  \frac{m_i}{384 \pi^3} \frac{g^4}{4} \, \int_0^{r_X^{max}} \, d r_X \, \lambda^{1/2}(1,r_X,r_j) \, \left( \sum_f (v_{W,f}^2 + a_{W,f}^2) N_c(f) \right) \nonumber\\
&&  \cdot \, \Big[ \left( |{\cal O}^{L}_{ji}|^2+|{\cal O}^{R}_{ji}|^2 \right)  \left( ( 1-r_j)^2 + (1+r_j-2r_X)r_X \right) - 12 Re[ {\cal O}^{L}_{ji} {\cal O}^{R*}_{ji} ] \, r_X \sqrt{r_j} \Big] \nonumber\\
&& \cdot \, \frac{1}{\left( (r_X-r_W)^2 + r_W r_{\Gamma_W} \right)},
\eea
where the maximum range of the integration is 
\beq
r_X^{max} \= (1-|\sqrt{r_j}|)^2.
\eeq
We sum over all final states into which off-shell gauge bosons can decay. The relevant coupling factors are
\beq
\left( \sum_f (v_{Z,f}^2 + a_{Z,f}^2) N_c(f) \right) \= 1.562, \qquad \left( \sum_f (v_{W,f}^2 + a_{W,f}^2) N_c(f) \right) \= 4.5.
\eeq
These are present in previous literatures \cite{Baer:1998bj,Martin:2000eq}, but here we express them in a form closest to the two-body decay widths into on-shell $W,Z$ in the limit of vanishing SM fermion masses. Our numerical results agree with previous ones. In the limit of small mass-splitting $\Delta \equiv (1 - m_j/m_i) \ll 1$, the three-body width is approximated as
\beq
\Gamma(\chi_i^0 \to \chi_j^0 Z^*) \, \sim \, \frac{g^4}{100 \pi^3} \, \frac{m_i^5}{m_Z^4} \, \Delta^5
\eeq
where the $\Delta^5$-dependence is notable. The three-body decay widths via $h$ is smaller as $h$ is narrower and heavier than $W,Z$. 

Two-body decays of heavier Higgsinos into pions and lighter Higgsinos are important when the mass splitting is smaller than about 1 GeV \cite{Chen:1996ap}. In most of parameter space we consider, tree-level mass splitting is greater than a few GeV, thus we assume that three-body decays are more important. 

The three-body decays between neutralinos may compete with the loop-induced two-body decays into photons~\cite{Haber:1988px,Ambrosanio:1996gz}. As a rough estimate, the loop-induced magnetic moment operator mediated via a charged Higgsino and a $W$ boson is $\sim g^2 e / 16\pi^2$. With an extra momentum factor from the magnetic operator, the loop-induced width scales with $\Delta^3$. After all, this decay mode is typically smaller than three-body modes in our parameter space. But more dedicated comparison will be interesting beyond our work.

%%%%%
\section{The derivation of the width ratio in the equivalence limit} \label{sec:app-gold}

When $m_{\chi_i}-m_{\chi_j} \gg m_h, m_Z$, the couplings of the $Z$ boson are understood as being inherited from Goldstone's. We first generalize neutral scalar couplings
\bea
D^L_{\phi^0 i j} &=& \frac{1}{\sqrt{2}}\left( k_{u\phi^0}^* U^*_{i1} V^*_{j2} + k_{d\phi^0}^* U^*_{i2} V^*_{j1} \right), \\
D^{\prime L}_{\phi^0 i j} &=& \frac{1}{2}( N_{j2}^* - t_W N_{j1}^* )( - k_{u\phi^0}^* N_{i4} + k_{d\phi^0}^* N^*_{i3}  ) + \Delta D^\prime_{\phi^0 ij}+ (i \leftrightarrow j )  \label{eq:nnphi}\\
\Delta D^\prime_{\phi^0 ij} &=& \frac{c_H \mu }{\sqrt{2} g v_{PQ}} N_{i5}^* \left(  k_{u \phi^0} N_{j3} + k_{d\phi^0} N_{j4} \right) 
\eea 
where
\bea
k_{u\phi^0} &=& ( c_\alpha, s_\alpha, i c_\beta, i s_\beta ), \qquad k_{d\phi^0} \= ( -s_\alpha, c_\alpha, i s_\beta, -i c_\beta ) \nonumber\\
&=& (-s_\beta, c_\beta, i c_\beta, i s_\beta), \qquad \quad \, \= (-c_\beta, -s_\beta, is_\beta, -ic_\beta )
\eea
for $h^0, H^0, A^0, G^0$ in order. The Goldstone coupling can be re-expressed in terms of $Z$ boson couplings \cite{Dreiner:2008tw}
\bea
D^{\prime L}_{G ij} &=& \frac{i}{2}( N_{j2}^* - t_W N_{j1}^*)( c_\beta N_{i3}^* + s_\beta N_{i4} ) + \frac{i c_H \mu}{\sqrt{2} g v_{PQ}} N_{ia} ( c_\beta N_{j4} - s_\beta N_{j3} ) \+  (i \leftrightarrow j) \\
&=& -i \frac{\sqrt{2}}{vg} \left( m_{\chi_i^0} {\cal O}^{\prime \prime L}_{ij} - m_{\chi_j^0} {\cal O}^{\prime \prime R}_{ij} \right) \= \frac{-i}{m_Z c_W} {\cal O}^{\prime \prime L}_{ij} ( m_{\chi_i^0}  + m_{\chi_j^0} )
\eea
where we have used ${\cal O}^{\prime \prime R}_{ij} = -{\cal O}^{\prime \prime L}_{ji} = -{\cal O}^{\prime \prime L}_{ij}$. Note that the relation is exact and is not modified by the existence of axino contributions, $\Delta D^\prime$. Furthermore, the Goldstone couplings are enhanced by $m_i/m_Z$ compared to $Z$ couplings which is analogous to the $y_t/g$ enhancement of the Goldstone couplings to top quarks.

The decay width into $G^0$ obtained from \Eq{eq:dec-neu-h} can now be expressed in the equivalence limit (as mentioned, the sign is $-2 \sqrt{r_j}$ for $G^0$ compared to $+2 \sqrt{r_j}$ for $h$)
\bea
\Gamma(\chi_i^0 \to \chi_j^0 G^0) &=& \frac{g^2 \,  m_{\chi_i^0}}{16\pi} \lambda^{1/2}\cdot  \left| D^{\prime L}_{Gij} \right|^2  \left( 1+r_j-r_h - 2 \sqrt{r_j} \right) \nonumber\\
&\simeq&  \frac{g^2 m_{\chi_i^0}}{16\pi c_W^2} \lambda^{1/2} \cdot \left| {\cal O}^{\prime \prime L}_{ij} \right|^2 \frac{m_i^2}{m_Z^2} \left( 1 +  {\cal O}(r) \right).
\eea
The same limiting formula is obtained from the partial width decay into $Z$, \Eq{eq:dec-neu-Z},
\beq
\Gamma(\chi_i^0 \to \chi_j^0 Z ) \, \simeq \, \frac{g^2 \,  m_{\chi_i^0}}{16\pi c_W^2} \lambda^{1/2} \cdot \left| {\cal O}^{\prime \prime L}_{ij} \right|^2 \, \frac{1}{r_Z} \, \simeq \, \frac{g^2 \,  m_{\chi_i^0}}{16\pi } \lambda^{1/2} \cdot \left| D^{\prime L}_{Gij} \right|^2 \left( 1- 2 \sqrt{r_j} \right)
\eeq
Meanwhile, the decay into the $h$ boson, \Eq{eq:dec-neu-h}, is approximated as
\beq
\Gamma(\chi_i^0 \to \chi_j^0 h ) \, \simeq \, \frac{g^2 \,  m_{\chi_i^0}}{16\pi} \lambda^{1/2} \cdot  \left| D^{\prime L}_{hij} \right|^2   \left( 1+ 2 \sqrt{r_j} \right)
\eeq
Now we usefully express partial width ratio in terms of the scalar coupling ratio
\beq
\frac{\Gamma(\chi_i^0 \to \chi_j^0 Z)}{\Gamma(\chi_i^0 \to \chi_j^0 h)} \, \simeq \, \frac{ | D^{\prime L}_{Gij}|^2 \, (1-2\sqrt{r_j})}{ | D^{\prime L}_{hij}|^2 \, (1+2 \sqrt{r_j})} .
\eeq
This relation is true regardless of the existence of axinos.

Likewise, we derive the width ratio for chargino decays in the equivalence limit. In this limit,
\beq
\Gamma(\chi_i^+ \to \chi_j^+ G) \, \simeq \, \frac{g^2 m_i}{32\pi} \lambda^{1/2} \Big( |D^L_{Gij}|^2 + |D^L_{Gji}|^2 + 4Re[ D^L_{Gij} D^L_{Gji} ] \sqrt{r_j} \Big)
\eeq
To approximate the full decay width into the $Z$ in terms of Goldstone couplings, we again use the identity \cite{Dreiner:2008tw}
\beq
D^L_{Gij} \= \frac{-i}{m_Z c_W} \left( m_i {\cal O}^{\prime L}_{ij} - m_j {\cal O}^{\prime R}_{ij} \right)
\eeq
Using the simplifying relation ${\cal O}^{\prime L}_{ij} = {\cal O}^{\prime L *}_{ji}$ holding for $i \ne j$ (and similarly for ${\cal O}^{\prime R}$), we obtain
\beq
|D^L_{Gij}|^ + |D^L_{Gji}|^2 \= \frac{1}{m_Z^2 c_W^2} \Big(  (m_i^2+m_j^2) ( |{\cal O}^{\prime L}_{ij}|^2 + |{\cal O}^{\prime R}_{ij}|^2) - 4 m_i m_j {\rm Re}[{\cal O}^{\prime L}_{ij} {\cal O}^{\prime R *}_{ij} ] \Big)
\ceq
Re[ D^L_{Gij} D^L_{Gji} ] \= \frac{-1}{m_Z^2 c_W^2} \Big( -(m_i^2+m_j^2) {\rm Re}[{\cal O}^{\prime L}_{ij} {\cal O}^{\prime R *}_{ij}] + m_i m_j (|{\cal O}^{\prime L}_{ij}|^2 + |{\cal O}^{\prime R}_{ij}|^2) \Big).
\eeq
By solving these for $|{\cal O}^{\prime L}_{ij}|^2+|{\cal O}^{\prime R}_{ij}|^2$ and ${\rm Re}[{\cal O}^{\prime L}_{ij} {\cal O}^{\prime R *}_{ij}]$, one can rewrite the partial width into the $Z$ boson as
\beq
\Gamma(\chi_i^+ \to \chi_j^+ Z ) \, \simeq \, \frac{g^2 m_i}{32 \pi} \lambda^{1/2} \Big( |D^L_{Gij}|^2 + |D^L_{Gji}|^2 + 4 \sqrt{r_j} Re[ D^L_{Gij} D^L_{Gji} ] \Big)
\eeq
which agrees with the Goldstone partial width above. Now we express partial width ratio in terms of scalar couplings
\beq
\frac{\Gamma(\chi_i^+ \to \chi_j^+ Z)}{\Gamma(\chi_i^+ \to \chi_j^+ h)} \, \simeq \, \frac{ |D^L_{Gij}|^2 + |D^L_{Gji}|^2 + 4Re[ D^L_{Gij} D^L_{Gji}] \sqrt{r_j} }{ |D^L_{hij}|^2 + |D^L_{hji}|^2 + 4Re[ D^L_{hij} D^L_{hji}] \sqrt{r_j} }.
\eeq

%%%%%%%%%%

%%%
\end{document}